# NEW PHYSICS SEARCHES AT LEP

S. BRAIBANT , G. GIACOMELLI

*University of Bologna and INFN, Bologna, Italy*

M. HAUSCHILD

*CERN, Geneva, Switzerland*

A short summary is made of several searches for new physics performed at the highest LEP2 energies.

## 1. Introduction

The Standard Model (SM) of Particle Physics explains quite well all available experimental results; the theory had very precise confirmations from the LEP high precision measurements [1]. On the other hand the theory contains too many free parameters, there is a proliferation of quarks and leptons and it seems unthinkable that there is no further unification with the strong interaction. Furthermore the presence of neutrino oscillations and neutrino masses indicate possible new physics beyond the SM [2].

In the following are summarized some recent highest LEP2 energies up to $E_{cm}$=209 GeV. They concern: excited leptons, leptoquarks, single top production, multi-photon events with large missing energy, stable or long-lived charged particles and a test of non-commutative QED.

## 2. Excited Leptons

The existence of 3 families of quarks and leptons is a strong motivation for substructures. In composite models, quarks, leptons and gauge bosons are composite with an associated energy scale $\Lambda$. These theories predict the existence of excited fermions F* with the same electroweak gauge couplings to the vector bosons as the SM

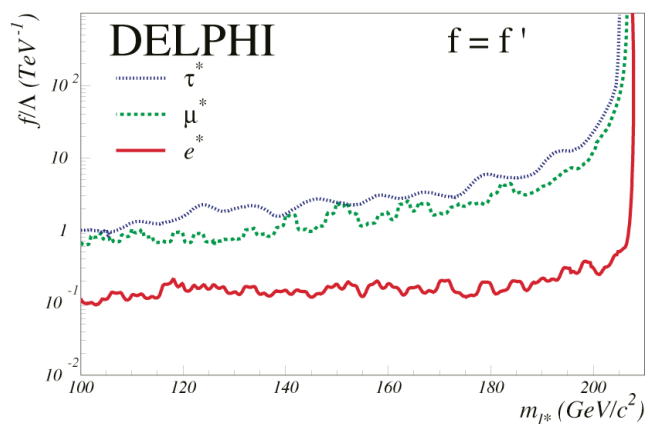

Fig. 1. DELPHI results on the single production of excited charged leptons assuming f=f'. The lines are 95% CL limits on f/$\Lambda$ versus excited lepton mass. These results will soon be combined with those of the other LEP groups in order to searches for new physics, performed at the     fermions. Excited fermions could be produced in pairs, $e^+e^- \rightarrow F^*\bar{F}^*$, or singly, $e^+e^- \rightarrow F^+\bar{F}^*$. Assuming photonic decays, $F^* \rightarrow f\gamma$, the final states involve two leptons and two photons. For excited neutrino-antineutrino pairs, the final state involves two photons plus missing energy/momentum.

Experimental results on this subject were presented by the DELPHI, L3 and OPAL experiments using LEP2 data up to 209 GeV [3]. Aside from one possible candidate event in OPAL, no other events were observed: thus there is no evidence for excited fermions with masses $< 202$ GeV, see Fig. 1 [3].

## 3. Leptoquarks

Leptoquarks (LQs) are particles that carry both lepton and baryon numbers. They are predicted in models which try to explain formally the

LEP2 energies [4]. The leptoquarks are assumed to be produced via couplings to the $\gamma/Z^0$. For a given search channel, only LQ decays involving a single generation have been considered. Aside from one possible candidate, no evidence for LQ pair production was observed, and lower limits on masses for scalar and vector LQs were calculated, improving the mass limits by 10-25 GeV, see Table 1. Notice the limits on the third family, which is not easily accessible at other accelerators.

## 4. Single top quark production

Fig. 2 illustrates the $e^+e^- \rightarrow \gamma/Z \rightarrow t\,\bar{c}$ process via Flavor Changing Neutral Currents, which are absent at tree level in the SM (even at the one-loop level they are severely suppressed). The study of single top production may thus be symmetry between quarks and leptons. They could be produced in pairs, and each LQ decays into lepton+quark.

Table 1. Summary of OPAL lower limits on leptoquarks masses.

| LQ | $Q_{e.m.}$ | $\beta$ | $1^{st}$ gen. | $2^{nd}$ gen. | $3^{rd}$ gen. |
|---|---|---|---|---|---|
| $S_0$ | $-1/3$ | [0.5,1] | 69(**) | 79(**) | 45(*) |
| $\tilde{S}_0$ | $-4/3$ | 1 | 99 | 100 | 98 |
| $S_1$ | $+2/3$ | 0 | 97 | 97 | 97 |
| | $-1/3$ | 0.5 | 69 | 79 | 45(*) |
| | $-4/3$ | 1 | 100 | 101 | 99 |
| $S_{1/2}$ | $-2/3$ | [0,1] | 94(**) | 94(**) | 93(**) |
| | $-5/3$ | 1 | 100 | 100 | 98 |
| $\tilde{S}_{1/2}$ | $+1/3$ | 0 | 89 | 89 | 89 |
| | $-2/3$ | 1 | 97 | 99 | 96 |
| $V_0$ | $-2/3$ | [0.5,1] | 99(**) | 99(**) | 97(**) |
| $\tilde{V}_0$ | $-5/3$ | 1 | 102 | 102 | 101 |
| $V_1$ | $+1/3$ | 0 | 101 | 101 | 101 |
| | $-2/3$ | 0.5 | 99 | 99 | 97 |
| | $-5/3$ | 1 | 102 | 102 | 101 |
| $V_{1/2}$ | $-1/3$ | [0,1] | 99(**) | 99(**) | 98(**) |
| | $-4/3$ | 1 | 102 | 102 | 101 |
| $\tilde{V}_{1/2}$ | $+2/3$ | 0 | 99 | 99 | 99 |
| | $-1/3$ | 1 | 101 | 101 | 99 |

(*): LEP1. (**): Minimum value $\forall$ $\beta = B.R.$ ($LQ \rightarrow l^-q$)

OPAL extended the search up the highest

used as a probe for new physics. In fact several extensions of the SM could enhance single top production. DELPHI [5] found no evidence for single top production and placed exclusion limits in the coupling parameters, $k_\gamma < 0.36$, $k_Z < 0.39$ for $m_t = 174$ GeV.

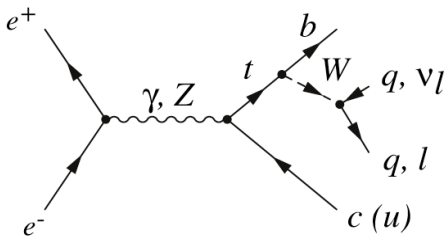

Fig. 2. Feynman diagram for single top production.

## 5. Search for acoplanar photons

Events with a final state consisting of two photons and large missing transverse energy

## 6. Stable and long-lived charged particles

OPAL made a search for stable or long-lived massive particles of electric charge Q/e=1 or fractional charges of 2/3, 4/3, measuring the ionization energy loss in their tracking chambers [7]. These particles are assumed to be pair produced in $e^+e^-$ collisions, not to interact strongly and to have lifetimes longer than $10^{-6}$ s. No evidence for the production of these particles was observed. Model-independent upper limits on the production cross sections were established for the production of scalar and spin-1/2 particles with charge ±1. This may imply mass lower limits of about 98 GeV for scalar muons and scalar taus. Long-lived charged heavy leptons and charginos are have been observed by OPAL [6] at the highest LEP2 energies. The results have been compared with the predictions of SM processes: no evidence was found for new physics contributions. This allowed to place new upper limits on various possible final states, like pair production of excited neutrinos, neutralino production and also, in the context of specific supersymmetric models, on light gravitinos.

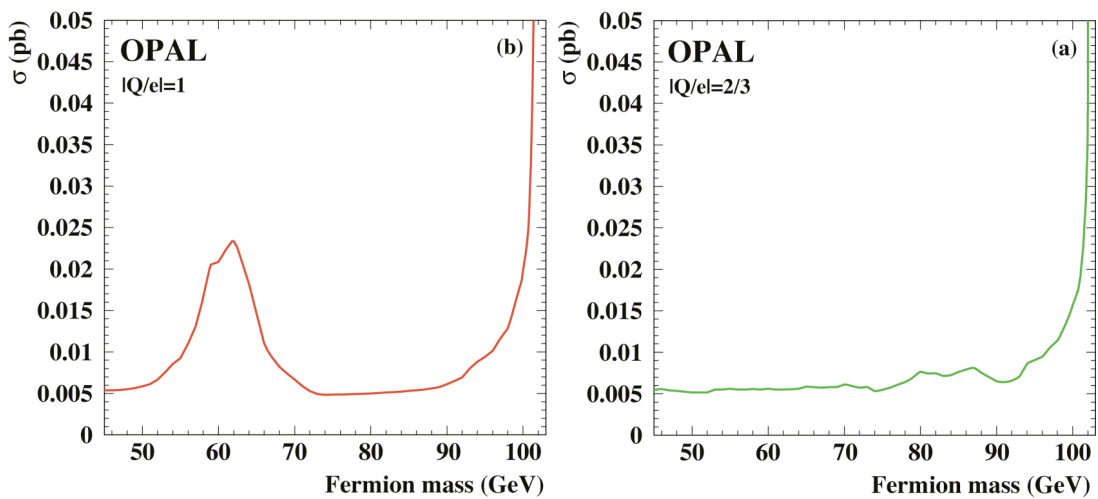

Fig. 3. 95% CL upper limits for the pair-production of long-lived charged fermions with |Q/e| = 1, 2/3.

excluded for masses below 102 GeV. Cross section limits are also established for fractional charge particles, see Fig.3.

## 7. Test of non commutative QED

OPAL made the first experimental study of non-commutative QED using the purely electrodynamics process $e^+e^- \rightarrow \gamma\gamma$ [8]. Non-commutative QED would lead to deviations from the SM depending on a new energy scale Λ and a unique direction in space. The predictions of a tree level calculation evaluated for the specific orientation of the OPAL detector have been compared with the measured data. The distributions of the polar and azimuth photon angles are used to extract limits on the energy scale Λ: there are no deviations from the SM prediction and they place the limit Λ>167 GeV, as shown in Fig. 4.

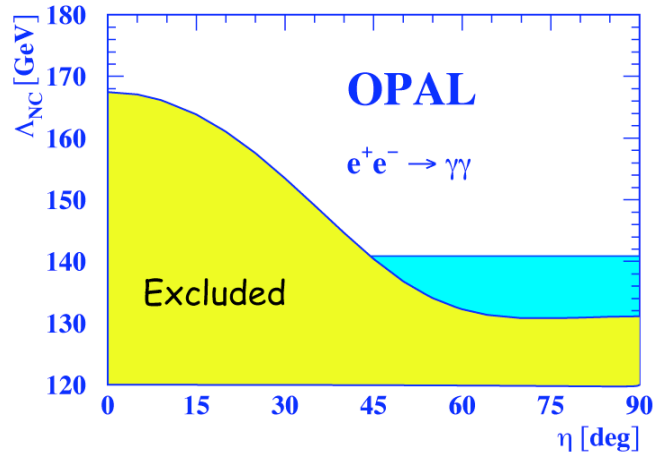

Fig. 4. Limits on the energy scale $\Lambda_{NC}$ as a function of the angle η.

## 8. Conclusions

LEP provided many data samples offering a great variety of experimental probes for the investigation of new phenomena and searches for new physics. No indications for physics beyond the SM were observed. New limits have been established in the frameworks of different models. The LEP Exotica Working Group is combining the results obtained by the different experiments

We thank many colleagues who helped to prepare this presentation. We apologize for not having been able to present all results.